# Astrophysical Data Analytics based on Neural Gas Models, using the Classification of Globular Clusters as Playground

Giuseppe Angora

Department of Physics "E. Pancini", University Federico II, Via Cinthia 6, 80126 Napoli, Italy
gius.angora@gmail.com

Massimo Brescia, Giuseppe Riccio

INAF Astronomical Observatory of Capodimonte, Via Moiariello 16, 80131 Napoli, Italy

Stefano Cavuoti, Maurizio Paolillo

Department of Physics "E. Pancini", University Federico II, Via Cinthia 6, 80126 Napoli, Italy

Thomas H. Puzia

Institute of Astrophysics, Pontificia Universidad Católica de Chile, Av. Vicuña Mackenna 4860, Macul, Santiago, Chile

**Abstract.** In Astrophysics, the identification of candidate Globular Clusters through deep, wide-field, single band HST images, is a typical data analytics problem, where methods based on Machine Learning have revealed a high efficiency and reliability, demonstrating the capability to improve the traditional approaches. Here we experimented some variants of the known Neural Gas model, exploring both supervised and unsupervised paradigms of Machine Learning, on the classification of Globular Clusters, extracted from the NGC1399 HST data. Main focus of this work was to use a well-tested playground to scientifically validate such kind of models for further extended experiments in astrophysics and using other standard Machine Learning methods (for instance Random Forest and Multi Layer Perceptron neural network) for a comparison of performances in terms of purity and completeness.

## 1 Introduction

The current and incoming astronomical synoptic surveys require efficient and automatic data analytics solutions to cope with the explosion of scientific data amounts to be processed and analyzed. This scenario, quite similar to other scientific and social contexts, pushed all communities involved in data-driven disciplines to explore data mining techniques and methodologies, most of which connected to the Machine Learning (hereafter ML) paradigms, i.e. supervised/unsupervised self-adaptive learning and parameter space optimization [3], [6], [7] .

Following this premise, this paper is focused on the investigation about the use of a particular kind of ML methods, known as Neural Gas (NG) models [21], to solve classification problems within the astrophysical context, characterized by a complex multi-dimensional parameter space.

In order to scientifically validate such models, we decided to approach a typical astrophysical playground, already solved with ML methods [8], [11] and to use in parallel other two ML techniques, chosen among the most standard, respectively, Random Forest [5] and Multi Layer Perceptron Neural Network [23], as comparison baseline.

The astrophysical case is related to the identification of Globular Clusters (GCs) in the galaxy NGC1399 using single band photometric data obtained through observations with the Hubble Space Telescope (HST) [8], [25], [27].

The physical identification and characterization of a Globular Cluster (GC) in external galaxies is considered important for a variety of astrophysical problems, from the dynamical evolution of binary systems, to the analysis of star clusters, galaxies and cosmological phenomena [27].

Here, the capability of ML methods to learn and recognize peculiar classes of objects, in a complex and noising parameter space and by learning the hidden correlation among object's parameters, has been demonstrated particularly suitable in the problem of GC classification [8]. In fact, multi-band wide-field photometric data (colours and luminosities) are usually required to recognize GCs within external galaxies, due to the high risk of contamination of background galaxies, which appear indistinguishable from galaxies located few Mpc away, when observed by ground-based instruments. Furthermore, in order to minimize the contamination, high-resolution space-borne data are also required, since they are able to provide particular physical and structural features (such as concentration, core radius, etc.), thus improving the GC classification performance [25].

In [8] we demonstrated the capability of ML methods to classify GCs using only single band images from Hubble Space Telescope with a classification accuracy of 98.3%, a completeness of 97.8% and only 1.6% of residual contamination. Thus confirming that ML methods may yield low contamination by minimizing the observing requirements and extending the investigation to the outskirts of nearby galaxies.

These results gave us an optimal playground where to train NG models and to validate their potential to solve classification problems characterized by complex data with a noising parameter space.

The paper is structured as follows: in Sect. 2 we describe the data used to test of the various methods. In Sect. 3 we provide a short methodological and technical description of the models. In Sect. 4 we describe the experiments and results about the parameter space analysis and classification experiments, while in Sect. 5 we discuss the results and draw our conclusions.

## 2 The Astrophysical Playground

As introduced, the HST single band data used are very suitable to investigate the classification of GCs. They, in fact, are deep and complete in terms of wide-field coverage, i.e. able to sample the GC population, to ensure a high S/N ratio required to measure structural parameters [10]. Furthermore, they provide the possibility to study the overall properties of the GC populations, which usually may differ from those of the central region of a galaxy.

With such data we intend to verify that Neural Gas based models could be able to identify GCs with low contamination even with single band photometric information. Throughout the confirmation of such behavior, we are confident that these models could solve other astrophysical problems as well as in other data-driven problem contexts.

### 2.1 The data

The data used in the described experiment consist of wide field single band HST observations of the giant elliptical NGC1399 galaxy, located in the core of the Fornax cluster [27]. Due to its distance (D=20.130 Mpc, see [13]), it is considered an optimal case where to cover a large fraction of its GC system with a restricted number of observations. This dataset was used by [25] to study the GC-LMXB connection and the structural properties of the GC population. The optical data were taken with the HST Advanced Camera for Surveys, in the broad V band filter, with 2108 seconds of integration time for each field. The observations were arranged in a 3x3 ACS mosaic with a scale of 0.03 arcsec/pix, and combined into a single image using the MultiDrizzle routine [19]. The field of view of the ACS mosaic covers ~100 square arcmin (Figure 1), extending out to a projected galacto-centric distance of ~55 kpc.

The source catalog was generated using Sextractor [4], [2], by imposing a minimum area of 20 pixels: it contains 12915 sources and reaches 7σ detection at m_V=27.5, i.e. 4 mag below the GC luminosity function, thus allowing to sample the entire GC population (see [8] for details).

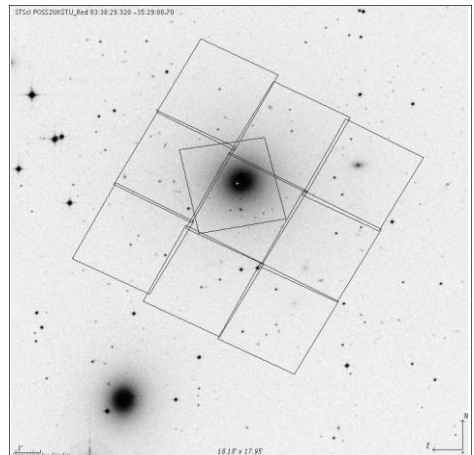

**Figure 1** The FoV covered by the HST/ACS mosaic in the broad V band.

The source subsample used to build our Knowledge Base (KB) to train the ML models, is composed by 2100 sources with 11 features (7 photometric and 4 morphological parameters).

Such parameter space includes three aperture magnitudes within 2, 6 and 20 pixels (*mag_aper1, mag_aper2, mag_aper3*), isophotal magnitude (*mag_iso*), kron radius (*kron_rad*), central surface brightness (*mu0*), FWHM (*fwhm_im*), and the four structural parameters, respectively, *ellipticity*, King's tidal, effective and core radii (*calr_t, calr_h, calr_c*). The target values of the KB required as ground truth for training and validation, i.e. the binary column indicating the source as GC or not GC, is provided through the typical selection based on multi-band magnitude and colour cuts. The original 2100 sources having a target assigned have been randomly shuffled and split into a training (70%) and a blind test set (30%).

## 3 The Machine Learning Models

In our work we tested three different variants of the Neural Gas model, using two additional machine learning methods, respectively feed-forward neural network and Random Forest, as comparison benchmarks. In the following all main features of these models are described.

### 3.1 Growing Neural Gas

Growing Neural Gas (GNG) is presented by [14] as a variant of the Neural Gas algorithm (introduced by

[21]), which combines the Competitive Hebbian Learning (CHL, [22]) with a vector quantization technique to achieve a learning that retains the topology of the dataset.

Vector quantization techniques [22] encode a data manifold, e.g. $V \subseteq \mathbb{R}^m$, using a finite set of reference vectors $w = w_1 \ldots w_N$, $w_i \in \mathbb{R}^m, i = 1 \ldots N$. Every data vector $v \in V$ is described by the best matching reference vector $w_{i(v)}$ for which the distortion error $d(v, w_{i(w)})$ is minimal. This procedure divides the manifold $V$ into a number of subregions: $V_i = v \in V : \|v - w_i\| \leq \|w - w_j\| \, \forall j$, called Voronoi polyhedra [24], within which each data vector $v$ is described by the corresponding reference vector $w_i$.

The Neural Gas network is a vector quantization model characterized by N neural units, each one associated to a reference vector, connected to each other. When an input is extracted, it induces a synaptic excitation detected by all the neurons in the graph and causes its adaptation. As shown in [21], the adaptation rule can be described as a "winner-takes-most" instead of "winner-takes-all" rule:

$$\Delta w_i = \varepsilon h_\lambda(v, w_i)(v - w_i), \; i = 1 \ldots N. \quad (1)$$

The step size $\varepsilon$ describes the overall extent of the adaptation. While $h_\lambda(v, w_i) = h_\lambda(k_i(v, w))$ is a function in which $k_i$ is the "neighborhood-ranking" of the reference vectors. Simultaneously, the first and second Best Matching Units (BMUs) develop connections between each other [21].

Each connection has an "age"; when the age of a connection exceeds a pre-specified lifetime T, it is removed [21]. Martinez's reasoning is interesting [22]: they demonstrate how the dynamics of neural units can be compared to a gaseous system. Let's define the density of vector reference at location $u$ through $\rho(u) = F_{BMU(u)}^{-1}$, where $F_{BMU}(u)$ is the volume of Voronoi polyhedra. Hence, $\rho(u)$ is a step function on each Voronoi polyhedra, but we can still imagine that their volumes change slowly from one polyhedra to the next, with $\rho(u)$ continuous. In this way, it is possible to derive an expression for the average change:

$$\langle \Delta w_i \rangle \propto \frac{1}{\rho^{1+2/m}} \left( \partial_u P(u) - \frac{2+m}{m} \frac{P}{\rho} \partial_u \rho(u) \right) \quad (2)$$

where $P(u)$ is the data point distribution.

The equation suggests the name Neural Gas: the average change of the reference vectors corresponds to a motion of particles in a potential $V(u) = -P(u)$. Superimposed on the gradient of this potential there is a force proportional to $-\partial_u \rho(u)$, which points toward the direction of the space where the particle density is low.

Main idea behind the GNG network is to successively add new units to an initially small network, by evaluating local statistical measures collected during previous adaptation steps [14]. Therefore, each neural unit in the graph has associated a local reconstruction error, updated for the BMU at each iteration (i.e. each time an input is extracted): $\Delta error_{BMU} = \|w_{BMU} - v\|$.

Unlike the Neural Gas network, in the GNG the synaptic excitation is limited to the receptive fields related to the Best Matching Unit and its topological neighbors: $\Delta w_i = \varepsilon_i(v - w_i)$, $i \in (BMU, n), \forall n \in neighbours(BMU)$. It is no longer necessary to calculate the ranking for all neural units, but it is sufficient to determine the first and the second BMU.

The increment of the number of units is performed periodically: during the adaptation steps the error accumulation allows to identify the regions in the input space where the signal mapping causes major errors. Therefore, to reduce this error, new units are inserted in such regions [14].

An elimination mechanism is also provided: once the connections, whose age is greater than a certain threshold, have been removed, if their connected units remain isolated (i.e. without emanating edges), those units are removed [14].

### 3.2 GNG with Radial Basis Function

Fritzke describes an incremental Radial Basis Function (RBF) network suitable for classification and regression problems [14].

The network can be figured out as a standard RBF network [9], with a GNG algorithm as embedded clustering method, used to handle the hidden layer.

Each unit of this hybrid model (hereafter GNGRBF) is a single perceptron with an associated reference vector and a standard deviation. For a given input-output pair $(v, y), v \in \mathbb{R}^n, y \in \mathbb{R}^m$, the activation of the i-th unit is described by $D_i(v) = e^{-\frac{\|v - w_i\|}{\sigma_i^2}}$.

Each of the single perceptron computes a weighted sum of the activations: $O_i = \sum_j w_{ij} D_j(v), i = 1 \ldots m$.

The adaptation rule applies to both reference vectors forming the hidden layer and the RBF weights. For the first, the adaptation rule is the same of the updating rule for the GNG network, while for the weights:

$$\Delta w_{ij} = \eta D_j(y_i - O_i), i = 1 \ldots m, j \in N \quad (3)$$

Similarly to the GNG network, new units are inserted where the prediction error is high, updating only the Best Matching Unit at each iteration: $\Delta error_{BMU} = \sum_{i=1}^{m} y_i - O_i$.

### 3.3 Supervised Growing Neural Gas

The Supervised Growing Neural Gas (SGNG) algorithm is a modification of the GNG algorithm that uses class labels of data to guide the partitioning of data into optimal clusters [15], [20]. Each of the initial neurons is labelled with a unique class label. To reduce the class impurity inside the cluster, the original learning rule (1) is reformulated by considering the case where the BMU belongs or not to the same class of the neuron whose reference vector is the closest to the current input. Depending on such situation the SGNG learning rule is expressed alternatively as:

$$\begin{cases} \Delta w_n = -\varepsilon \frac{v-w_n}{\|v-w_n\|} \quad or \\ \Delta w_n = +\varepsilon \frac{v-w_n}{\|v-w_n\|} + repulsion(sn,n) \end{cases} \quad (4)$$

where $sn$ is the nearest class neuron and $repulsion()$ is a function specifically introduced to maintain neurons sufficiently distant one each other. For the neuron which is topologically close to the neuron $sn$, the rule intends to increase the clustering accuracy [20]. The insertion mechanism has to reduce not only the intra-distances between data in a cluster, but also the impurity of the cluster. Each unit has associated two kinds of error: an aggregated and a class error. A new neuron is inserted close to the neuron having a highest class error accumulated, while the label is the same as the neuron label with the greater aggregated error.

### 3.4 Multi Layer Perceptron

The Multi Layer Perceptron (MLP) architecture is one of the most typical feed-forward neural networks [23]. The term feed-forward is used to identify basic behavior of such neural models, in which the impulse is propagated always in the same direction, e.g. from neuron input layer towards output layer, through one or more hidden layers (the network brain), by combining the sum of weights associated to all neurons.

As easy to understand, the neurons are organized in layers, with proper own role. The input signal, simply propagated throughout the neurons of the input layer, is used to stimulate next hidden and output neuron layers. The output of each neuron is obtained by means of an activation function, applied to the weighted sum of its inputs.

The weights adaptation is obtained by the Logistic Regression rule [17], by estimating the gradient of the cost function, the latter being equal to the logarithm of the likelihood function between the target and the prediction of the model. In this work, our implementation of the MLP is based on the public library Theano [1].

### 3.5 Random Forest

Random Forest (RF) is one of the most widely known machine learning ensemble methods [5], since it uses a random subset of candidate data features to build an ensemble of decision trees. Our implementation makes use of the public library scikit-learn [26]. This method has been chosen mainly because it provides for each input feature a score of importance (rank) measured in terms of its informative contribution percentage to the classification results. From the architectural point of view, a RF is a collection (forest) of tree-structured classifiers $h(x, \theta_k)$, where the $\theta_k$ are independent, identically distributed random vectors and each tree casts a unit vote for the most popular class at input. Moreover, a fundamental property of the RF is the intrinsic absence of training overfitting [5].

## 4 The experiments

The five models previously introduced have been applied to the dataset described in Sec. 2.1 and their performances have been compared to verify the capability of NG models to solve particularly complex classification problems, like the astrophysical identification of GCs from single-band observed data.

### 4.1 The Classification Statistical Estimators

In order to evaluate the performances of the selected classifiers, we decided to use three among the classical and widely used statistical estimators, respectively, average efficiency, purity, completeness and F1-score, which can be directly derived from the confusion matrix [28], showed in Figure 2. The *average efficiency* (also known as accuracy, hereafter *AE*), is the ratio between the sum of correctly classified objects on both classes (true positives for both classes, hereafter tp) and the total amount of objects in the test set. The *purity* (als known as precision, hereafter *pur*) of a class measures the ratio between the correctly classified objects and the sum of all objects assigned to that class (i.e. tp/[tp+fp], where fp indicates the false positives). While the *completeness* (also known as recall, hereafter *comp*) of a class is the ratio tp/[tp+fn], where fn is the number of false negatives of that class. The quantity tp+fn corresponds to the total amount of objects belonging to that class. The *F1-score* is a statistical test that considers both the purity and completeness of the test to compute the score (i.e. 2[pur*comp]/[pur+comp]).

By definition, the dual quantity of the purity is the *contamination*, another important measure which indicates the amount of misclassified objects for each class.

| Confusion Matrix | Predicted Class GC | Predicted Class notGC |
|---|---|---|
| **True class GC** | tp | fn |
| **True class notGC** | fp | tn |

**Figure 2** The confusion matrix used to estimate the classification statistics. Columns indicate the class objects as predicted by the classifier, while rows are referred to the true objects of the classes. Main diagonal terms contain the number of correctly classified for the two classes, while fp counts the false positives and fn the false negatives of the GC class.

In statistical terms, it is well known the classical tradeoff between purity and completeness in any classification problem, particularly accentuated in astrophysical problems [12]. In the specific case of the GC identification, from the astrophysical point of view, we were mostly interested to the purity, i.e. to ensure the highest level of true GCs correctly identified by the classifiers [8]. However, within the comparison experiments described in this work, our main goal was to evaluate the performances of the classifiers mostly related to the best tradeoff between purity and completeness.

### 4.2 Analysis of the Data Parameter Space

Before to perform the classification experiments, we preliminarily investigated the parameter space, defined by the 11 features defined in Sec. 2.1, identifying each object within the KB dataset of 2100 objects. Main goal of this phase was to measure the importance of any feature, i.e. its relevance in terms of informative contribution to the solution of the problem. In the ML context, this analysis is usually called *feature selection* [16]. Its main role is to identify the most relevant features of the parameter space, trying to minimize the impact of the well known problem of the *curse of dimensionality*, i.e. the fact that ML models exhibit a decrease of performance accuracy when the number of features is significantly higher than optimal [18]. This problem is mainly addressed to cases with a huge amount of data and dimensions. However, its effects may also impact contexts with a limited amount of data and parameter space dimension.

The Random Forest model resulted particularly suitable for such analysis, since it is intrinsically able to provide a feature importance ranking during the training phase. The feature importance of the parameter space, representing the dataset used in this work, is shown in Figure 3.

From the astrophysical point of view, this ranking is in accordance with the physics of the problem. In fact, as expected, among the five most important features there are the four magnitudes, i.e. the photometric log-scale measures of the observed object's photonic flux through different apertures of the detector. Furthermore, almost all photometric features resulted as the most relevant. Finally, by looking at the Figure 3, there is an interesting gap between the first six and the last five features, whose cumulative contribution is just ~11% of the total. Finally, a very weak joined contribution (~3%) is carried by the two worst features (*kron_rad* and *calr_c*), which can be considered as the most noising/redundant features for the problem domain.

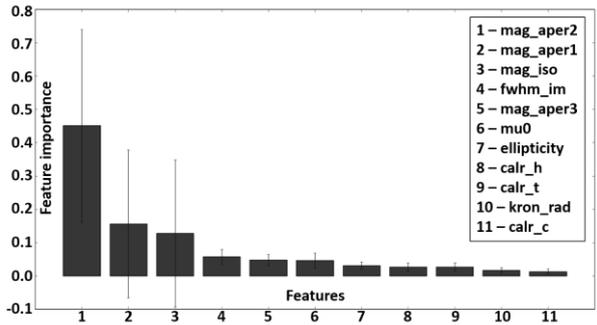

**Figure 3** The feature importance ranking obtained by the Random Forest on the 11-feature domain of the input dataset during training (see Sec. 2.1 for details). The blue vertical lines report the importance estimation error bars.

Based on such considerations, the analysis of the parameter space provides a list of most interesting classification experiments to be performed with the selected five ML models. This list is reported in Table 1.

The experiment E1 is useful to verify the efficiency by considering the four magnitudes.

The experiment E2 is based on the direct evaluation of the best group of features as derived from the importance results.

The classification efficiency of the full photometric subset of features is evaluated through the experiment E3.

Finally, the experiment E4 is performed to verify the results by removing only the two worst features.

**Table 1** List of selected experiments, based on the analysis of the parameter space. The third column reports the identifiers of the included features, according to the importance ranking (see legend in Figure 3).

| EXP ID | # features | included features |
|--------|------------|-------------------|
| E1 | 4 | 1,2,3,5 |
| E2 | 6 | 1,2,3,4,5,6 |
| E3 | 7 | 1,2,3,4,5,6,10 |
| E4 | 9 | 1,2,3,4,5,6,7,8,9 |

### 4.3 The Classification Experiments

Following the results of the parameter space analysis, the original domain of features has been reduced, by varying the number and types of included features. Therefore, the classification experiments have been performed on the dataset, described in Sec. 2.1, composed by 2100 objects and represented by a parameter space with up to a maximum of 9 features (Table 1).

**Table 2** Statistical analysis of the classification performances obtained by the five ML models on the blind test set for the four selected experiments. All quantities are expressed in percentage and related to average efficiency (*AE*), purity for each class (*purGC, purNotGC*), completeness for each class (*compGC, compNotGC*) and the F1-score for GC class. The contamination is the dual value of the purity.

| ID | Estimator | RF % | MLP % | SGNG % | GNGRBF % | GNG % |
|---|---|---|---|---|---|---|
| E1 | *AE* | 88.9 | 84.4 | 88.1 | 88.1 | 88.4 |
| | *purGC* | 85.9 | 80.1 | 89.7 | 85.4 | 83.7 |
| | *compGC* | 87.3 | 82.6 | 80.3 | 85.7 | 89.2 |
| | *F1-scoreGC* | 86.6 | 81.3 | 84.7 | 85.5 | 86.4 |
| | *purNotGC* | 91.0 | 87.6 | 87.2 | 90.0 | 92.1 |
| | *compNotGC* | 89.7 | 85.6 | 93.0 | 89.6 | 88.1 |
| E2 | *AE* | 89.0 | 85.1 | 87.3 | 88.3 | 83.2 |
| | *purGC* | 84.9 | 77.0 | 81.0 | 82.9 | 74.0 |
| | *compGC* | 89.2 | 90.7 | 90.3 | 90.0 | 91.1 |
| | *F1-scoreGC* | 87.0 | 83.3 | 85.4 | 86.3 | 81.7 |
| | *purNotGC* | 92.2 | 92.6 | 92.7 | 92.6 | 92.6 |
| | *compNotGC* | 89.0 | 85.6 | 85.7 | 87.4 | 80.0 |
| E3 | *AE* | 89.0 | 83.2 | 85.1 | 89.2 | 86.8 |
| | *purGC* | 85.2 | 77.2 | 80.0 | 86.0 | 84.1 |
| | *compGC* | 88.8 | 83.8 | 84.9 | 88.0 | 83.8 |
| | *F1-scoreGC* | 87.0 | 80.4 | 82.4 | 87.0 | 83.9 |
| | *purNotGC* | 91.9 | 88.0 | 89.0 | 91.5 | 88.7 |
| | *compNotGC* | 89.9 | 83.2 | 85.1 | 89.8 | 88.4 |
| E4 | *AE* | 89.5 | 86.0 | 88.1 | 88.7 | 83.8 |
| | *purGC* | 85.3 | 82.5 | 84.1 | 83.8 | 78.3 |
| | *compGC* | 90.0 | 83.8 | 87.6 | 90.0 | 83.8 |
| | *F1-scoreGC* | 87.6 | 83.1 | 85.8 | 86.8 | 81.0 |
| | *purNotGC* | 92.7 | 88.6 | 91.1 | 92.6 | 88.1 |
| | *compNotGC* | 89.1 | 87.5 | 88.1 | 88.2 | 84.1 |

The dataset has been randomly shuffled and split into a training set of 1470 objects (70% of the whole KB) and a blind test set of 630 objects (the residual 30% of the KB).

These datasets have been used to train and test the selected five ML classifiers. The analysis of results, reported in Table 2, has been performed on the blind test set, in terms of the statistical estimators defined in Sec. 4.2.

## 5 Discussion and Conclusions

As already underlined, main goal of this work is the validation of NG models as efficient classifiers in noising and multi-dimensional problems, with performances at least comparable to other ML methods, considered "traditional" in terms of their use in such kind of problems.

By looking at Table 2 and focusing on the statistics for the three NG models, it is evident that their result is able to identify GCs from other background objects, reaching a satisfying tradeoff between purity and completeness in all experiments and for both classes. The occurrence of statistical fluctuations is mostly due to the different parameter space used in the four experiments. Nevertheless, none of the three NG models overcome the others in terms of the measured statistics.

If we compare the NG models with the two additional ML methods (Random Forest and MLP neural network), their performances appears almost the same. This implies that NG methods show classification capabilities fully comparable to other ML methods.

Another interesting aspect is the analysis of the degree of coherence among the NG models in terms of commonalities within classified objects. Table 3 reports the percentages of common predictions for the objects correctly classified by considering, respectively both and single classes. On average, the three NG models are in agreement among them for about 80% of the objects correctly classified.

**Table 3** Statistics for the three NG models related to the common predictions of the correctly classified objects. Second column is referred to both classes, while the third and fourth columns report, respectively, the statistics for single classes.

| EXP ID | GC+notGC % | GC % | notGC % |
|---|---|---|---|
| E1 | 86.0 | 85.4 | 86.9 |
| E2 | 79.8 | 79.8 | 79.8 |
| E3 | 81.1 | 82.5 | 79.2 |
| E4 | 77.8 | 77.4 | 78.4 |

This is also confirmed by looking at the Figure 4, where the tabular results of Table 3 are showed through the Venn diagrams, reporting also more details about their classification commonalities.

Finally, from the computational efficiency point of view, the NG models have theoretically a higher complexity than Random Forest and neural networks. But, since they are based on a dynamic evolution of the internal structure, their complexity strongly depends on the nature of the problem and its parameter space.

Nevertheless, all the presented ML models have a variable architectural attitude to be compliant with the parallel computing paradigms. Besides the embarrassingly parallel architecture of the Random Forest, the use of optimized libraries, like Theano [1], make also models like MLP highly efficient. From this point of view NG models have a high potentiality to be parallelized. By optimizing GNG, the GNGRBF would automatically benefit, since both share the same search space, except for the RBF training additional cost. In practice, the hidden layer of the supervised network behaves just like a GNG network whose neurons act as inputs for the RBF network. Consequently, with the same number of iterations, the GNGRBF network performs a major number of operations.

On the other hand, the SGNG network is similar to the GNG network, although characterized by a neural insertion mechanism over a long period, thus avoiding too rapid changes in the number of neurons and excessive oscillations of reference vectors. Therefore, on average, the SGNG network computational costs are higher than the models based on the standard Neural Gas mechanism.

In conclusion, although a more intensive test campaign on these models is still ongoing, we can assert that Neural Gas based models are very promising as problem-solving methods, also in presence of complex and multi-dimensional classification and clustering problems, especially if preceded by an accurate analysis and optimization of the parameter space within the problem domain.

## Acknowledgements

MB acknowledges the PRIN-INAF 2014 *Glittering kaleidoscopes in the sky: the multifaceted nature and role of Galaxy Clusters*, and the PRIN-MIUR 2015 *Cosmology and Fundamental Physics: illuminating the Dark Universe with Euclid.*
MB, GL and MP acknowledge the H2020-MSCA-ITN-2016 SUNDIAL (*SUrvey Network for Deep Imaging Analysis and Learning*), financed within the Call H2020-EU.1.3.1.MB acknowledges the PRIN-INAF 2014 *Glittering kaleidoscopes in the sky: the multifaceted nature and role of Galaxy Clusters*, and the PRIN-MIUR 2015 *Cosmology and Fundamental Physics: illuminating the Dark Universe with Euclid.*
MB, GL and MP acknowledge the H2020-MSCA-ITN-2016 SUNDIAL (*SUrvey Network for Deep Imaging Analysis and Learning*), financed within the Call H2020-EU.1.3.1.

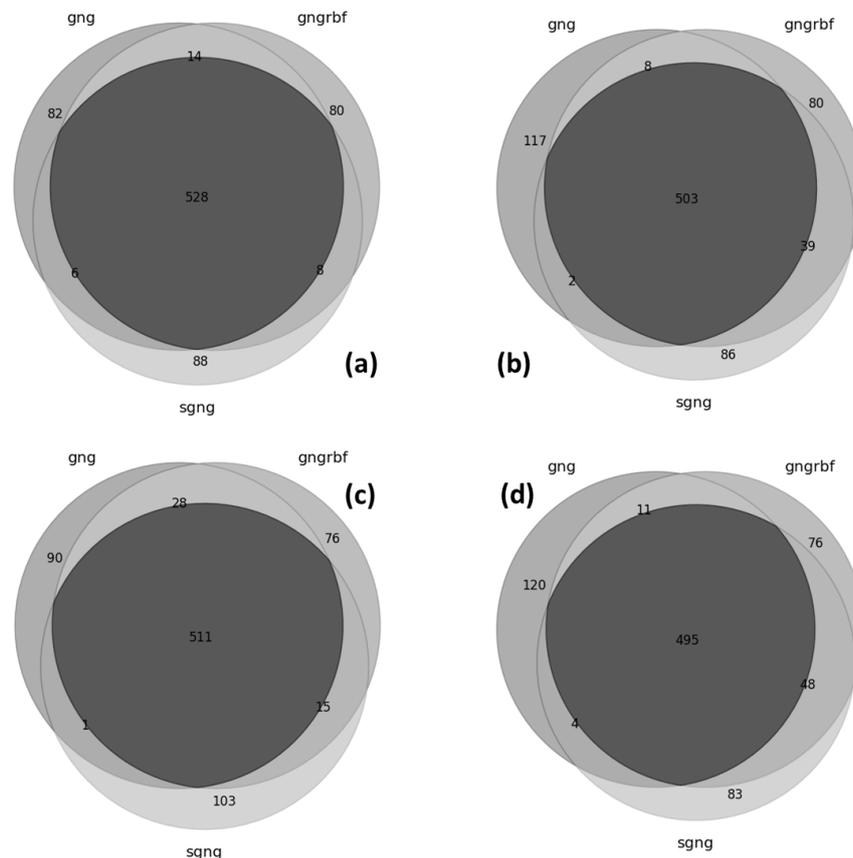

**Figure 4** The Venn diagram related to the prediction of all (both GCs and not GCs) correctly classified objects performed by the three Neural Gas based models (GNG, GNGRBF and SGNG) for the experiments, respectively, E1 (a), E2 (b), E3 (c) and E4 (d). The intersection areas (dark grey in the middle) show the objects classified in the same way by different models. Internal numbers indicate the amount of objects correctly classified for each sub-region.